 \documentclass[symmetry,article,accept,moreauthors,pdftex,usenames,dvipsnames,a4paper]{Definitions/mdpi} 

\firstpage{1} 
\pubvolume{14}
\issuenum{2}
\articlenumber{265}
\pubyear{2022}
\copyrightyear{2022}
\externaleditor{Academic Editors: {Vitaly Bornyakov, Konstantin Klimenko, Natalia Kolomoyets, Roman Rogalyov and Roman Zhokhov}} 
\datereceived{21 December 2021} 
\dateaccepted{22 January 2022} 
\datepublished{29 January 2022} 
\hreflink{https://doi.org/10.3390/\linebreak sym14020265} 
\pdfoutput=1

\newcommand{\cleverrefpackage}{
	\RequirePackage[capitalise]{cleveref}
}

\ifdef{\AddToHook}{
	\AddToHook{begindocument}{ 
		\cleverrefpackage
	}
}{
	\cleverrefpackage
}

\usepackage{graphicx}
\usepackage{subcaption}
\usepackage{mathtools}
\usepackage{amsmath}
\usepackage{xfrac}
\usepackage{xcolor}
\usepackage{dsfont}
\usepackage{booktabs}
\usepackage{bbold}
\usepackage[utf8]{inputenc}
\usepackage[ngerman, english]{babel}
\DeclareMathOperator{\tr}{tr}

\newcommand{\xt}[1]{#1}
\newcommand{\x}[1]{\textbf{#1}}
\renewcommand{\arraystretch}{1.5}
\newcommand{\Ns}{\ensuremath{N_{s}}}
\newcommand{\Nt}{\ensuremath{N_{t}}}
\newcommand{\Ntc}{\ensuremath{N_{t,c}}}
\newcommand{\Nf}{\ensuremath{N_{f}}}

\newcommand{\I}{\ensuremath{\mathds{1}}}
\newcommand{\ii}{\ensuremath{\mathrm{i}}}
\newcommand{\R}{\ensuremath{\mathbb{R}}}
\newcommand{\Z}{\ensuremath{\mathbb{Z}}}

\newcommand{\e}{\ensuremath{\mathrm{e}}}
\newcommand{\seff}{S_{\text{eff}}}

\newcommand{\seffL}{\seff^{L}}
\newcommand{\seffR}{\seff^{R}}

\newcommand{\seffI}{S_{\text{eff,I}}}
\newcommand{\kron}[2]{\ensuremath{\delta_{#1,#2}}}
\newcommand{\muff}{\ensuremath{\mu_{45}}}
\newcommand{\muR}{\ensuremath{\mu_{R}}}
\newcommand{\muL}{\ensuremath{\mu_{L}}}
\newcommand{\muI}{\ensuremath{\mu_{\text{I}}}}
\newcommand{\Qfour}{\ensuremath{Q^{(4)}}}
\newcommand{\QI}{\ensuremath{Q_{\text{I}}}}
\DeclareMathOperator{\Det}{Det}

\usepackage{xstring}

\newcommand{\gtapprox}{\raisebox{-0.5ex}{$\,\stackrel{>}{\scriptstyle\sim}\,$}}

\providecommand{\Rcite}[1]{%
	\begingroup
	\def\tempx{0}%
	\StrCount{#1}{,}[\tempx]%
	\ifnum\tempx > 0 
	Refs.~%
	\else
	Ref.~%
	\fi
	\endgroup
	\cite{#1}%
}

\Title{
 Inhomogeneous Phases in the Chirally Imbalanced $2+1$-Dimensional Gross-Neveu Model and Their Absence in the Continuum Limit
}

\TitleCitation{ Inhomogeneous Phases in the Chirally Imbalanced $2+ \linebreak 1$-Dimensional Gross-Neveu Model and Their Absence in the Continuum Limit}

\Author{Laurin Pannullo $^{1,}$*\orcidA{}, Marc Wagner $^{1,2}$\orcidB{} and Marc Winstel $^{1,}$*\orcidC{}}

\AuthorNames{Laurin Pannullo, Marc Wagner and Marc Winstel}

\AuthorCitation{Pannullo, L.; Wagner, M.; Winstel, M.}

\address{%
$^{1}$ \quad Institut für Theoretische Physik, Goethe Universität Frankfurt am Main, Max-von-Laue-Straße 1, \mbox{60438 Frankfurt, Germany}
\\
$^{2}$ \quad Helmholtz Research Academy Hesse for FAIR,
Campus Riedberg, Max-von-Laue-Straße 12
, \mbox{60438 Frankfurt, Germany}}

\corres{Correspondence: pannullo@itp.uni-frankfurt.de (L.P.); winstel@itp.uni-frankfurt.de (M.W.)}

\abstract{We studied the $\mu$-$\muff$-$T$ phase diagram of the $2+1$-dimensional Gross-Neveu model, where $\mu$ denotes the ordinary chemical potential, $\muff$ the chiral chemical potential and $T$ the temperature. We use the mean-field approximation and two different lattice regularizations with naive chiral fermions.
An inhomogeneous phase at finite lattice spacing was found for one of the two regularizations.
Our results suggest that there is no inhomogeneous phase in the continuum limit.
We showed that a chiral chemical potential is equivalent to an isospin chemical potential. Thus, all results presented in this work can also be interpreted in the context of isospin imbalance.}


\begin{document}

\section{Introduction}

The Gross-Neveu (GN) model describes a theory of $\Nf$ fermion flavors with a quartic interaction. It is a rather simple model commonly used to explore and describe the spontaneous breaking of chiral symmetry \cite{Gross:1974jv} in the $\mu$-$T$ plane, where $\mu$ denotes the chemical potential and $T$ the temperature. 
In the limit $\Nf \rightarrow \infty$ (corresponding to the mean-field approximation or, equivalently, the  neglect of bosonic quantum fluctuations) the \mbox{$1+1$-dimensional} GN model exhibits three phases: a symmetric phase (with a vanishing chiral condensate), a homogeneous symmetry-broken phase (with a non-zero, but spatially constant condensate) and an inhomogeneous phase, where the chiral condensate is an oscillating function of space \cite{Thies:2003kk,Schnetz:2004vr,Thies:2006ti}.
The phase diagrams of the GN model and related theories were also investigated at finite $\Nf$, i.e., with bosonic quantum fluctuations included, using lattice Monte-Carlo simulations \cite{Cohen:1983nr,Karsch:1986hm,Lenz:2020bxk,Lenz:2020cuv,Lenz:2021kzo,Lenz:2021vdz, Horie:2021wnn} and the functional renormalization group (FRG)~\cite{Stoll:2021ori}. 

Inhomogeneous phases are not limited to the GN model, but were found in several models in the mean-field approximation in $1+1$ dimensions \cite{Basar:2009fg,Thies:2018qgx, Thies:2019ejd, Thies:2021shf} and in $3+1$ dimensions {\cite{Kutschera:1989yz,Kutschera:1990xk,Nakano:2004cd,Nickel:2009wj,Broniowski:2011ef,Carignano:2014jla,Heinz:2015lua}}. 
For a review, we refer to \Rcite{Buballa:2014tba}.
In recent works \cite{Pisarski:2020dnx, Pisarski:2020gkx, Pisarski:2021qof, Rennecke:2021ovl} it has been discussed that inhomogeneous phases might be related to so-called moat regimes, where the bosonic wave function renormalization $Z$ is negative. 
Such a regime has been found in an FRG study of the phase diagram of quantum chromodynamics (QCD) \cite{Fu:2019hdw}. 
A similar regime has been observed in the $1+1$-dimensional GN model (see Figure\ 9 of \Rcite{Koenigstein:2021llr}
, where the wave function renormalization $Z$ is plotted in the $\mu$-$T$ plane).
One finds that a negative $Z$ accompanies the instability of a homogeneous condensate with respect to inhomogeneous perturbations as a necessary, but not a sufficient condition.
The possible existence of a moat regime in the QCD phase diagram encourages us to improve our understanding of inhomogeneous condensation and related phenomena in QCD-inspired models.

Recently, the existence of inhomogeneous phases was also explored in the $2+1$-dimensional GN model in the mean-field approximation \cite{Winstel:2019zfn,Narayanan:2020uqt,Buballa:2020nsi}. 
Such $2+1$-dimensional four-fermion theories are of interest both in high energy physics \cite{Hands:1992ck,Hands:2001cs,Gies:2010st,Scherer:2012nn,Narayanan:2020ahe} and in condensed matter physics \cite{MacKenzie:1992qr,Tesanovic:2002zz,Wen:2002zz,Rantner:2002zz,Ohsaku:2003rq,Kalinkin:2003uw,Hands:2008id,Ziegler:2020zkq}, but also to study conceptual questions, e.g., renormalizability in the $1/N$ expansion or in a perturbative approach \cite{Hands:1991py,Wellegehausen:2017goy,Hands:2020itv,Hands:2021mrg}.
Hence, confirming the existence of an inhomogeneous phase in such a model could have a significant impact.
Early seminal studies of the $\mu$-$T$ phase diagram of the $2+1$-dimensional GN model \cite{Klimenko:1987gi, Rosenstein:1988dj}  reported a second-order phase transition between the symmetric and the homogeneous symmetry-broken phase at finite $T$ and $\mu$ and a first-order phase transition at $T = 0$. However, in these studies only a homogeneous order parameter was considered.
In our recent publication \Rcite{Buballa:2020nsi} we studied the existence of an inhomogeneous phase in the $2+1$-dimensional GN model within the mean-field approximation. 
Our main findings were that an inhomogeneous phase is present at finite regulator and for certain regularization schemes (a Pauli-Villars cutoff and a specific lattice discretization), but it disappears when the regulator is removed, as previously observed in \Rcite{Narayanan:2020ahe}. 

In this work we continue our investigations from \Rcite{Buballa:2020nsi} by extending the $2+1$-dimensional GN model with a chiral chemical potential. We studied its phase diagram, where our main focus was on possibly existing inhomogeneous phases. While the GN model might be too simple to realistically describe the effect of chiral imbalance on QCD, it might still improve our conceptual understanding of inhomogeneous condensation in the presence of chiral imbalance, which is an important problem.
A difference in the densities of left- and right-handed quarks is relevant in physical systems such as heavy-ion collisions \cite{Kharzeev:2001ev,Kharzeev:2004ey} or compact stars \cite{Charbonneau:2009ax,Ohnishi:2014uea}.
The impact of chiral imbalance on chiral symmetry breaking has been studied (e.g., in $1+1$-dimensional models) \cite{Thies:2018qgx}, where it had no influence on the existence of the inhomogeneous chiral spiral, and in $3+1$-dimensional models \cite{Khunjua:2018dbm, Chernodub:2020yaf,Khunjua:2021oxf}, where only homogeneous order parameters were considered. 
A chirally imbalanced $2+1$-dimensional GN model, extended by a quartic difermion interaction, was explored in \Rcite{Ebert:2016ygm, Zhukovsky:2017hzo} with the aim to clarify the competition of homogeneous fermion--fermion condensation and homogeneous chiral condensation. 
In recent two-color and three-color QCD studies \cite{Braguta:2015zta,Braguta:2015owi} a chiral chemical potential was investigated and found to increase the chiral transition temperature. This result is supported by a Nambu-Jona-Lasinio (NJL) model study \cite{Braguta:2016aov}.

This paper is structured as follows.
We start in \cref{theory} by discussing the theoretical basics of the GN model in $2+1$ dimensions including details on the underlying chiral symmetry.
We also add a chiral chemical potential to the model and show the equivalence of chiral imbalance and isospin imbalance.
In \cref{lattice} we discretize the effective action of the model using lattice field theory. 
\cref{results} is the main part of our paper, where numerical results are presented and discussed. Finally, we conclude in \cref{conclusion}. Preliminary results from this project were presented at a recent conference \cite{Winstel:2021yok}.


\section{Theoretical Basics \label{theory}} 


\subsection{Action and Partition Function}

The action of the GN model in $2+1$ dimensions with $\Nf$ fermion flavors is
\begin{align}
	\label{eq:fermi_action}
	S[\bar{\psi},\psi] = \int d^3x \, \bigg(\sum_{n=1}^{\Nf} \bar{\psi}_n \Big(\gamma_\nu \partial_\nu + \gamma_0 \mu\Big) \psi_n - \frac{g^2}{2} \bigg(\sum_{n=1}^{\Nf} \bar{\psi}_n \psi_n\bigg)^2\bigg) , 
\end{align}
where $\psi_n$ represents $\Nf$ massless fermion fields; $\mu$ is the chemical potential; and $g^2$ is the coupling of the four-fermion interaction. $\int d^3x = \int_0^{1/T} dx_0 \, \int_{\R^2} d^2x$ with $d^2x = dx_1 \, dx_2$ and $T$ denoting the temperature given by the inverse extent of the periodic temporal direction of Euclidean space-time.

The action (\ref{eq:fermi_action}) is equivalent to
\begin{align}
	\label{eq:Sbosonized}
	S[\bar{\psi},\psi,\sigma] = \int d^3x \, \bigg(\frac{\Nf}{2 \lambda} \sigma^2 + \sum_{n=1}^{\Nf} \bar{\psi}_n Q[\mu,\sigma] \psi_n\bigg) ,
\end{align}
where $\sigma$ is a scalar boson field; $\lambda = \Nf g^2$ is the rescaled coupling; and
\begin{align}
	\label{eq:Q}
	Q[\mu,\sigma] = \gamma_\nu \partial_\nu + \gamma_0 \mu + \sigma
\end{align}
is the Dirac operator. Integration over the fermion fields leads to the so-called effective action and the corresponding partition function
\begin{align}
	\label{eq:S_eff}
	\seff[\sigma] = \Nf \bigg(\frac{1}{2 \lambda} \int d^3x \, \sigma^2 - \ln \Det Q[\mu,\sigma]\bigg) , \quad Z = \int D\sigma \, e^{-\seff[\sigma]} .
\end{align}

One can show that $\langle \sigma(x) \rangle$ is related to the condensate $\langle \bar{\psi}_n(x) \psi_n(x) \rangle$ according to
\begin{align}
	\label{eq:sigm_eqofmotion}
	\langle \sigma(x) \rangle = -\frac{\lambda}{\Nf} \langle \bar{\psi}_n(x) \psi_n(x) \rangle .
\end{align}

As in previous studies of chiral inhomogeneous phases, 
we restricted the dependence of $\sigma$ to the spatial coordinates, i.e., $\sigma = \sigma(\x{x})$. 
With this restriction $\Det Q$ is real, which is shown in the Appendix of Ref.\ \cite{Buballa:2020nsi}.
For even $\Nf$, the effective action $\seff[\sigma]$ is then real. (Our numerical calculations of the determinant showed that $\Det Q$ is exclusively positive, i.e., the effective action $\seff[\sigma]$ is real for all values of $\Nf$).

Since $\seff[\sigma] \propto \Nf$, the limit $\Nf \rightarrow \infty$ reduces the relevant configurations in the partition function (\ref{eq:S_eff}) to the global minima of $\seff[\sigma]$. Thus, the computation of a path integral is reduced to an optimization problem. In the case of degenerate global minima, spontaneous symmetry breaking selects one of these minima. Consequently, an expectation value $\langle O(\sigma) \rangle$ is identical to the value of $O$ evaluated at the corresponding global minimum, i.e., $\langle O(\sigma) \rangle \rightarrow O(\sigma)$.  In particular, $\langle \sigma \rangle \rightarrow \sigma$. For the remainder of this paper we exclusively consider the limit $\Nf \rightarrow \infty$.


\subsection{Representation of the Dirac Matrices and Chiral Symmetry}

Typically one uses either an irreducible $2 \times 2$ representation or a reducible $4 \times 4$ representation of the Dirac algebra for the $\gamma$ matrices appearing in the Dirac operator
\labelcref{eq:Q} (for details see, e.g., \Rcite{Pisarski:1984dj,Gies:2010st,Scherer:2012nn,Buballa:2020nsi}). In the case of an irreducible $2 \times 2$ representation, there is no symmetry, which can be interpreted as chiral symmetry, because it is impossible to define a matrix $\gamma_5$, which anticommutes with $\gamma_0$, $\gamma_1$ and $\gamma_2$. Therefore, a reducible $4\times4$ representation is more appropriate in our context, e.g.: 
\begin{align}
	& \gamma_0 = \tau_3 \otimes \tau_2 = \Bigg(\begin{array}{cc}
	+\tau_2 & 0 \\
	0 & -\tau_2
	\end{array}\Bigg) , \quad \gamma_1 = \tau_3 \otimes \tau_3 = \Bigg(\begin{array}{cc}
	+\tau_3 & 0 \\
	0 & -\tau_3
	\end{array}\Bigg) , \nonumber \\
	\label{eq:4comp_gamma}
	& \gamma_2 = \tau_3 \otimes \tau_1 = \Bigg(\begin{array}{cc}
	+\tau_1 & 0 \\
	0 & -\tau_1
	\end{array}\Bigg) ,
\end{align}
where $\tau_j$ denote the Pauli matrices. The three matrices $+\tau_1$, $+\tau_2$ and $+\tau_3$ as well as the three matrices $-\tau_1$, $-\tau_2$ and $-\tau_3$ form irreducible $2 \times 2$ representations, which are inequivalent. The corresponding upper two and lower two entries of the fermion fields $\psi_n$ can be interpreted as left-handed and right-handed components, respectively.

The Dirac operator \labelcref{eq:Q} is then block-diagonal,
\begin{equation}
	\label{eq:block_Q4_balanced}
	Q[\mu,\sigma] = \Qfour[\mu,\sigma] = \Bigg(\begin{array}{cc}
	Q^{(2)}[\mu,\sigma] & 0 \\
	0 & \tilde{Q}^{(2)}[\mu,\sigma]
	\end{array}\Bigg) ,
\end{equation}
where
\begin{align}
	\label{eq:Q2}
	Q^{(2)}[\mu,\sigma] & = +\tau_2 (\partial_0 + \mu) + \tau_3 \partial_1 + \tau_2 \partial_2 + \sigma , \\
	\label{eq:Qtilde2}
	\tilde{Q}^{(2)}[\mu,\sigma] & = -\tau_2 (\partial_0 + \mu) - \tau_3 \partial_1 - \tau_2 \partial_2 + \sigma
\end{align}
represent Dirac operators for left-handed and right-handed fermion fields $\psi_{n}^{L/R}$ (see also Equations\ (14) and\ (15) in \Rcite{Buballa:2020nsi}).
One can show that $\Det Q^{(2)}[\mu,\sigma]$ and $\Det \tilde{Q}^{(2)}[\mu,\sigma]$ are invariant under both $\mu \rightarrow -\mu$ and $\sigma \rightarrow -\sigma$. Using the latter one can show
\begin{equation}
	\label{eq: Q2_eq_Qtilde2}
	\Det Q^{(2)}[\mu,\sigma] = \Det \tilde{Q}^{(2)}[\mu,\sigma] .
\end{equation}

The action with $Q = \Qfour[\mu,\sigma]$ is invariant under the discrete chiral transformations. (For free fermions one can define continuous axial chiral symmetry transformations with both $\gamma_4$ and $\gamma_5$. The four-fermion interaction in Equation\ \labelcref{eq:fermi_action} breaks the corresponding symmetries explicitly, see, e.g., \Rcite{Buballa:2020nsi}).
\begin{align}
	\label{eq:gamma4_disc_sym}
	& \psi_n \rightarrow \gamma_4 \psi_n , \quad \bar{\psi}_n \rightarrow -\bar{\psi}_n \gamma_4 ,  \\
	\label{eq:gamma5_disc_sym}
	& \psi_n \rightarrow \gamma_5 \psi_n , \quad \bar{\psi}_n \rightarrow -\bar{\psi}_n \gamma_5
\end{align}
with
\begin{align}
	\label{eq:gamma4_gamma5}
	\gamma_4 = \tau_1 \otimes \I_2 = \Bigg(\begin{array}{cc}
	0 & +\I_2 \\
	+\I_2 & 0
	\end{array}\Bigg) , \quad \gamma_5 = -\tau_2 \otimes \I_2 = \Bigg(\begin{array}{cc}
	0 & +\ii\I_2 \\
	-\ii\I_2 & 0
	\end{array}\Bigg) .
\end{align}

Both $\gamma_4$ and $\gamma_5$ anticommute with $\gamma_0$, $\gamma_1$ and $\gamma_2$, thus fulfilling the necessary properties for generating an axial chiral transformation. The symmetries \labelcref{eq:gamma4_disc_sym} and \labelcref{eq:gamma5_disc_sym} are also present for the action \labelcref{eq:Sbosonized}, where the corresponding transformation of $\sigma$ is in both cases $\sigma \rightarrow -\sigma$. Thus, $\sigma$ is an order parameter for chiral symmetry breaking. (When using an irreducible $2 \times 2$ fermion representation, there is no chiral symmetry; however, $\sigma$ can still be interpreted as an order parameter for parity breaking).

In addition to the transformations \labelcref{eq:gamma4_disc_sym,eq:gamma5_disc_sym} the action is also invariant under the continuous vector chiral transformations
\begin{align}
	\label{eq:phase_vector_trafo}
	& \psi_n \rightarrow \e^{\ii \alpha^a T^a} \psi_n , \quad \bar{\psi}_n \rightarrow \bar{\psi}_n \e^{-\ii \alpha^a T^a} , \\
	\label{eq:vector_trafo}
	& \psi_n \rightarrow \e^{\ii \beta^a T^a \gamma_{45}} \psi_n ,\quad \bar{\psi}_n \rightarrow \bar{\psi}_n \e^{-\ii \beta^a T^a \gamma_{45}} ,
\end{align} 
where $T^a$ denotes the generators of $\mathrm{U}(\Nf)$ flavor rotations and
\begin{equation}
	\gamma_{45} = \ii \gamma_4 \gamma_5 = \tau_3 \otimes \I_2 =  \Bigg(\begin{array}{cc}
	+\I_2 & 0\\
	0 & - \I_2
	\end{array}\Bigg)
\end{equation}
(see also \Rcite{Pisarski:1984dj,Gies:2010st}).

The transformations \labelcref{eq:gamma4_disc_sym}, \labelcref{eq:gamma5_disc_sym} and \labelcref{eq:vector_trafo} are not independent. For example, \labelcref{eq:gamma4_disc_sym} can be written as combination of \labelcref{eq:gamma5_disc_sym} and \labelcref{eq:vector_trafo} with $\beta^a T^a = -\pi/2$,
\begin{align}
	\label{eq:double_trafo}
	\psi_n \rightarrow \e^{\ii (-\pi/2) \gamma_{45}} \gamma_{5} \psi_n = \gamma_4 \psi_n , \quad \bar{\psi}_n \rightarrow -\bar{\psi}_n \gamma_5 \e^{-\ii (-\pi/2) \gamma_{45}} = -\bar{\psi}_n \gamma_4 .
\end{align}

Thus, there is only one independent $\Z_2$ symmetry, i.e., the structure of chiral symmetry is $\mathrm{U}_{\I}(\Nf) \times \mathrm{U}_{\gamma_{45}}(\Nf) \times \Z_2$.

A chiral chemical potential $\muff$ can be introduced in a straightforward way by extending and replacing the Dirac operator in \labelcref{eq:Q} or equivalently \labelcref{eq:block_Q4_balanced} according to
\begin{eqnarray}
	\nonumber
	 & & \hspace{-0.7cm} Q[\mu,\sigma] = \Qfour[\mu,\sigma] \rightarrow Q[\mu,\muff,\sigma] = \Qfour[\mu,\muff,\sigma] = \gamma_\nu \partial_\nu + \gamma_0 \mu + \gamma_{45} \gamma_0 \muff + \sigma = \\
	\label{eq:block_Q4}
	& & = \Bigg(\begin{array}{cc}
	Q^{(2)}[\mu + \muff,\sigma] & 0 \\
	0 & \tilde{Q}^{(2)}[\mu - \muff,\sigma]
	\end{array}\Bigg) .
\end{eqnarray}
$\muff$ contributes to the chemical potentials of the left-handed (upper two) components and the right-handed (lower two) components with opposite sign, thus causing chiral imbalance.
We note that there are other possibilities to define chirality and chiral imbalance (see, e.g., \Rcite{Ebert:2016ygm, Zhukovsky:2017hzo} and \cref{conclusion}) differing from our definition, where left- and right-handed fermion fields $\psi_{n}^{L/R}$ are projected from the fermion fields as
\begin{equation}
		\psi_{n}^{L/R} = P^{L/R} \psi_n = \tfrac{1}{2}(\I_4 \pm \gamma_{45}) \psi_n,  
		\label{eq:projections}
\end{equation} 
with $P^{L/R}$ denoting the corresponding projectors.

As done for $\muff = 0$ in appendix~A of \Rcite{Buballa:2020nsi}, one can show that $\Det \Qfour[\mu,\muff,\sigma]$ is invariant under both $(\mu,\muff) \rightarrow (-\mu,-\muff)$ and $\sigma \rightarrow -\sigma$. Since
\begin{equation}
	\label{eq:iso_det}
	\Det \Qfour[\mu,\muff,\sigma] = \Det Q^{(2)}[\mu + \muff,\sigma] \Det \tilde{Q}^{(2)}[\mu - \muff,\sigma] ,
\end{equation}
$\Det \Qfour[\mu,\muff,\sigma]$ is also invariant under the exchange of the ordinary and the chiral chemical potential, $\mu \leftrightarrow \muff$. Clearly, $\seff[\sigma]$ as well as the phase diagram share this invariance. In \cref{results} we use this property to cross-check our numerical results. 

We note that the effective action can be written as the sum of a left-handed and a right-handed part,
\begin{align}
	\nonumber \seff[\sigma] & = \Nf \bigg(\frac{1}{2 \lambda} \int d^3x \, \sigma^2 - \ln \Det \Qfour[\mu,\muff,\sigma]\bigg) = \seffL[\sigma] + \seffR[\sigma] = \\
	\label{eq:seff_LR}
	& = \sum_{X=L,R} \underbrace{\Nf \bigg(\frac{1}{2 (2 \lambda)} \int d^3x \, \sigma^2 - \ln \Det Q^{(2)}[\mu_X,\sigma]\bigg)}_{= S_\text{eff}^X[\sigma]}
\end{align}
with $\muL = \mu + \muff$ and {$\muR = \mu - \muff$}. Of course, the two parts are not independent but coupled via $\sigma$.
Moreover, both parts are equivalent to the chirally balanced effective action (see Section II C.~ of \Rcite{Buballa:2020nsi}), i.e.,
\begin{align}
	\seffL[\sigma] = \frac{1}{2} \seff[\sigma]\bigg|_{\mu = \muL, \muff = 0} , \quad \seffR[\sigma] = \frac{1}{2} \seff[\sigma]\bigg|_{\mu = \muR, \muff = 0} .
\end{align}
This property will be useful when we discuss our numerical results in \cref{results}.


\subsection{Equivalence of Isospin and Chiral Imbalance}

In this subsection we consider an even number of fermion flavors $\Nf$, again in the 4-component reducible representation, and assign half of them (the ``$u$ flavors'') a chemical potential $\mu + \muI$, the other half (the ``$d$ flavors'') a chemical potential $\mu - \muI$. Clearly, $\muI$ generates an imbalance between the $u$ and the $d$ flavors and, thus, can be interpreted as isospin chemical potential.

The corresponding effective action is
\begin{align}
	\label{eq:S_effI}
	\seffI[\sigma] = \Nf \bigg(\frac{1}{2 \lambda} \int d^3x \, \sigma^2 - \frac{1}{2} \ln \Det \QI[\mu,\mu_I,\sigma]\bigg) ,
\end{align}
where the Dirac operator is an $8 \times 8$ matrix in spin and isospin space,
\begin{align}
	\label{eq:block_QI}
	\QI[\mu,\mu_I,\sigma] &= \gamma_\nu \partial_\nu + \gamma_0 \mu + \gamma_0 \tau_3 \muI + \sigma=
	\nonumber \\
	& = \bigg(\begin{array}{cc}
	\Qfour[\mu + \muI,0,\sigma] & 0 \\
	0 & \Qfour[\mu - \muI,0,\sigma]
	\end{array}\bigg) .
\end{align}

Using Equation (\ref{eq:iso_det}) one can show
\begin{align}
	\label{eq:QI_eq_Qch}
	\ln \Det \QI[\mu,\mu_I,\sigma] = 2 \ln \Det \Qfour[\mu,\muI,\sigma] .
\end{align}

Consequently, the effective action for the GN model with isospin imbalance, Equation  (\ref{eq:S_effI}), is identical to the the effective action for the GN model with chiral imbalance, Equation~(\ref{eq:seff_LR}), when identifying $\muI = \muff$. Thus, all numerical results presented in Equation (\ref{results}) can either be interpreted in the context of chiral imbalance or of isospin imbalance. We note that this equivalence of isospin and chiral imbalance is specific to the GN model in $2+1$ dimensions.


\section{Lattice Discretization \label{lattice}}

We used a lattice discretization of the effective action \labelcref{eq:seff_LR}, which was similar to the discretization discussed in Section IV of our previous work \cite{Buballa:2020nsi}
. The key difference is that we use the naive fermion discretization also in temporal and not only in the spatial directions. 

We considered a 3-dimensional space-time volume $\beta V$, where $\beta = 1/T$ was the inverse temperature and $V = L^2$ the quadratic spatial volume. The boundary conditions were periodic in the 2 spatial directions and periodic and antiperiodic in temporal direction for the fields $\sigma$ and $\psi_n, \bar{\psi}_n$, respectively. We used a cubic lattice with $\Nt \times \Ns^2$ lattice sites and lattice spacing $a$, i.e., $\beta = a \Nt$ and $L = a \Ns$. In the following, all dimensionful quantities are expressed in units of the lattice spacing, e.g.,\ $a \equiv 1$, $\beta \equiv \beta / a$. Because of the finite space-time volume, the $3$-dimensional momenta were quantized,
\begin{align}
	\xt{p} = (p_0,\x{p}) = 2 \pi \bigg(\frac{k_0 + \eta}{\Nt} , \frac{\x{k}}{\Ns}\bigg)
	\label{eq:3dMomenta}
\end{align}
with
\begin{align}	
	k_0 \in \bigg\{-\frac{\Nt}{2} , -\frac{\Nt}{2}+1 , \ldots , \frac{\Nt}{2}-1 \bigg\} \quad \text{and} \quad \x{k}_i \in \bigg\{-\frac{\Ns}{2} , -\frac{\Ns}{2}+1 , \ldots , \frac{\Ns}{2}-1\bigg\}\nonumber
\end{align}
and $\eta = 0, 1/2$, corresponding to periodic and antiperiodic boundary conditions in the temporal direction.

In our numerical implementation, the effective action and fields were treated in momentum space,
\begin{align}
	\label{eq:seff_momentum_disc}
	\frac{\seff[\sigma]}{\Nf} = \frac{\Nt \Ns^2}{2 \lambda} \sum_{\x{p}} \tilde{\sigma}^2(\x{p}) - \frac{1}{8} \ln \Det \tilde{Q}^{(4)}[\mu,\muff,\sigma] , 
\end{align}
where
\begin{align}
	\tilde{\sigma}(\x{p}) = \frac{1}{\Ns^2}\sum_{\x{x}} \sigma(\x{x}) \e^{\ii \x{x} \cdot \x{p}}
\end{align}
are the Fourier coefficients of the field $\sigma(\x{x})$.
\begin{align}
	\label{eq:diracop_disc}
	\tilde{Q}^{(4)}_{\xt{p},\xt{q}}[\mu,\muff,\sigma] = \Nt\Ns^2\bigg(&\ii \kron{\xt{p}}{\xt{q}} \sum_{\nu=0}^{2} \gamma_\nu \sin\Big(p_\nu - \kron{\nu}{0} \ii (\mu + \gamma_{45} \muff)\Big) +
	\nonumber \\
	&+ \kron{p_0}{q_0} \tilde{W}_2(\x{p}-\x{q}) \tilde{\sigma}(\x{p}-\x{q}) \bigg) 
\end{align}
is the Dirac operator in momentum space. On the lattice, this operator is a matrix with columns and rows labeled by the momenta $\xt{p}$ and $\xt{q}$, respectively. The $\sin$ for $\nu = 0$ contains the matrix $\gamma_{45}$ but can be simplified according to
\begin{align}
	\sin\Big(p_0 - \ii (\mu + \gamma_{45} \muff)\Big) = \sin(p_0 - \ii \mu) \cos(\ii \muff) - \gamma_{45} \cos(p_0 - \ii \mu) \sin(\ii \muff) .
\end{align}

The Dirac operator $\tilde{Q}^{(4)}_{\xt{p},\xt{q}}$ has eight regions of soft modes, where the dispersion relation is approximately linear, in the first Brioullin zone and where each region describes a fermion flavor. In the continuum limit, these fermion flavors do not interact with each other but with the scalar field $\sigma$ in the same manner (for details see \Rcite{Cohen:1983nr} and the Appendix of \Rcite{Lenz:2020bxk}).
Thus, in order to study $\Nf$ fermion fields on the lattice, where $\Nf$ is restricted to a multiple of $8$, one has to use $\Nf/8$ naive fermions in the discretization of the fermion bilinear in Equation (\ref{eq:Sbosonized}) resulting in the factor $1/8$ in Equation  (\ref{eq:seff_momentum_disc}) (compare Section IV A.~in \Rcite{Buballa:2020nsi}).
An appropriately chosen weight function $\tilde{W}_2(\x{p})$ was necessary to ensure the correct continuum limit (see \Rcite{Cohen:1983nr,Lenz:2020bxk,Buballa:2020nsi} for details). We investigated and compared two possible choices,
\begin{align}
	\label{eq:W'_p}
	& \tilde{W}_2(\x{p}) = \tilde{W}'_2(\x{p}) = \prod_{\nu=1,2} \tilde{W}'_1(p_\nu) , \quad \tilde{W}'_1(p_\nu) = \frac{\cos(p_\nu) + 1}{2} , \\
	\label{eq:W''_p}
	& \tilde{W}_2(\x{p}) = \tilde{W}''_2(\x{p}) = \prod_{\nu=1,2} \tilde{W}''_1(p_\nu) , \quad \tilde{W}''_1(p_\nu) = \Theta(\pi/2 - |p_\nu|)
\end{align}
with $\Theta$ denoting the Heaviside function.

Because of the restriction of $\sigma$ to the spatial coordinates, i.e.,\ $\sigma = \sigma(\x{x})$, the Dirac operator \labelcref{eq:diracop_disc} was block-diagonal with respect to $p_0$ and $q_0$. This simplified the computation of $\Det \tilde{Q}^{(4)}[\mu,\muff,\sigma]$ to the computation of $\Nt$ determinants of smaller matrices of size $4 \Ns^2 \times 4 \Ns^2$.


\section{Numerical Results}\label{results}

Using lattice field theory the phase diagram of the $2+1$-dimensional GN model with $\mu_{45} = 0$ was extensively explored in \Rcite{Winstel:2019zfn,Narayanan:2020uqt,Buballa:2020nsi}. There is a symmetric phase with $\sigma = 0$ at large $\mu$ or large $T$ and a homogeneous symmetry-broken phase with a constant $\sigma = \bar{\sigma}$ at small $\mu$ and small $T$. Moreover, at finite lattice spacing and for certain discretizations (e.g., $\tilde{W}_2 = \tilde{W}''_2$) there is additionally an inhomogeneous phase, where $\sigma(\x{x})$ is a varying function of the spatial coordinates.
However, this inhomogeneous phase shrinks, when decreasing the lattice spacing, and seems to vanish in the continuum limit.

The main focus of this paper is to investigate in particular the phase structure for $\muff \neq 0$ to clarify whether inhomogeneous phases exist. At first, we recalled that the effective action $\seff[\sigma]$ can be written as the sum of a left-handed part $\seffL[\sigma]$ and a right-handed part $\seffR[\sigma]$ with chemical potentials $\muL$ and $\muR$, respectively (see Equation (\ref{eq:seff_LR})). Moreover, each of the two parts was equivalent to the action of the chirally balanced GN model, which was investigated in detail in our previous work \cite{Buballa:2020nsi}. Thus, for $|\muL| > \mu_c(T)$ and $|\muR| > \mu_c(T)$ both $\seffL[\sigma]$ and $\seffR[\sigma]$ had their respective minima at $\sigma = 0$ ($\mu_c(T)$, denoting the location of the phase boundary of the symmetric phase at $\muff = 0$ and temperature $T$). (We ignored the existence of an ``inhomogeneous island'' or ``inhomogeneous continent'' at large chemical potentials where cutoff effects were particularly strong 
 \cite{Buballa:2020nsi}.).
Consequently, the minimum of $\seff[\sigma]$ also corresponded to $\sigma = 0$. In other words, from numerical results obtained in \Rcite{Buballa:2020nsi} at $\muff = 0$ we could conclude that chiral symmetry was restored in the chirally imbalanced GN model for $|\mu_L| > \mu_c(T)$ and $|\mu_R| > \mu_c(T)$. In the remaining regions of the $(\mu, \muff, T)$  space, $\seffL[\sigma]$ and $\seffR[\sigma]$ compete and the behavior of the condensate needed to be investigated numerically. 
We started doing that in \cref{sec:hom_res} by restricting our computations to a homogeneous condensate. After that, in \cref{sec:instability}, we carried out a stability analysis of the favored value of the homogeneous condensate with respect to inhomogeneous perturbations. Finally, in \cref{sec:min}, we performed numerical minimizations of the effective action, allowing arbitrary inhomogeneous modulations.

The lattice spacing $a$ was a function of the coupling $\lambda$. As explained in our previous \mbox{work \cite{Buballa:2020nsi}} we tuned $\lambda$ such that the temporal extent $\Ntc a$ corresponded to the inverse critical temperature $\beta_c = 1/T_c$, which separated at $\mu = \mu_{45} = 0$, the symmetric and homogeneous symmetry-broken phase. Then, at fixed $\lambda$, the temperature $T = 1/\Nt a$ could be changed in discrete steps by increasing or decreasing $\Nt$. A summary of the lattice parameters used to generate all following numerical results is given in \cref{tab:a_lambda}. We note that throughout this section dimensionful quantities are expressed in units of the vacuum expectation value of $\sigma$,
\begin{equation}
\sigma_0 = \sigma|_{\mu = 0, \mu_{45} = 0, T = 0} .
\end{equation} 

\vspace{-9pt}
\begin{table}[H]
	\caption{Lattice parameters ($\Ntc$: number of lattice sites in temporal direction corresponding to the critical temperature $T_c$; $\lambda$: coupling; $a$: lattice spacing; $\Ns$: number of lattice sites in each of the two spatial directions).} \label{tab:a_lambda} 
	\renewcommand{\arraystretch}{0.8}
	\setlength{\tabcolsep}{11.75mm}\begin{tabular}{cccc}
		\toprule
		\boldmath{$\Ntc$} & \boldmath{$\lambda/a$} & \boldmath{$a \sigma_0$} & \boldmath{$\Ns$} \\
		\midrule
		$4$ & $2.6040$ & $0.3649$ & $28$, $40$, $60$, $80$ \\
		$6$ & $2.3355$ & $0.2327$ & $60$, $100$, $120$ \\
		\bottomrule
	\end{tabular}

\end{table}


\subsection{Restriction to a Homogeneous Condensate \label{sec:hom_res}}

In the case of a homogeneous condensate, i.e., $\sigma = \bar{\sigma}$ or equivalently $\tilde{\sigma}(\x{p}) = \bar{\sigma} \kron{\x{p}}{0}$, the two lattice discretizations with $\tilde{W}'_2$ and $\tilde{W}''_2$ (Equations (\ref{eq:W'_p}) and (\ref{eq:W''_p})) were identical. The Dirac operator corresponded to a block-diagonal matrix with $\Nt \Ns^2$ blocks of size $4 \times 4$. Thus, the $\ln \Det \tilde{Q}^{(4)}[\mu,\muff,\sigma]$ term in the effective action \labelcref{eq:seff_momentum_disc} could be computed quite efficiently by summing over $\Nt \Ns^2$ determinants of $4 \times 4$ matrices. Moreover, the effective action at given $\mu$, $\muff$ and $T$ is a function of just a single variable $\bar{\sigma}$; hence, it could be minimized numerically in a straightforward and rather cheap way to obtain the physically preferred value of the homogeneous condensate.

\begin{figure}[H]
	\includegraphics[width=\columnwidth]{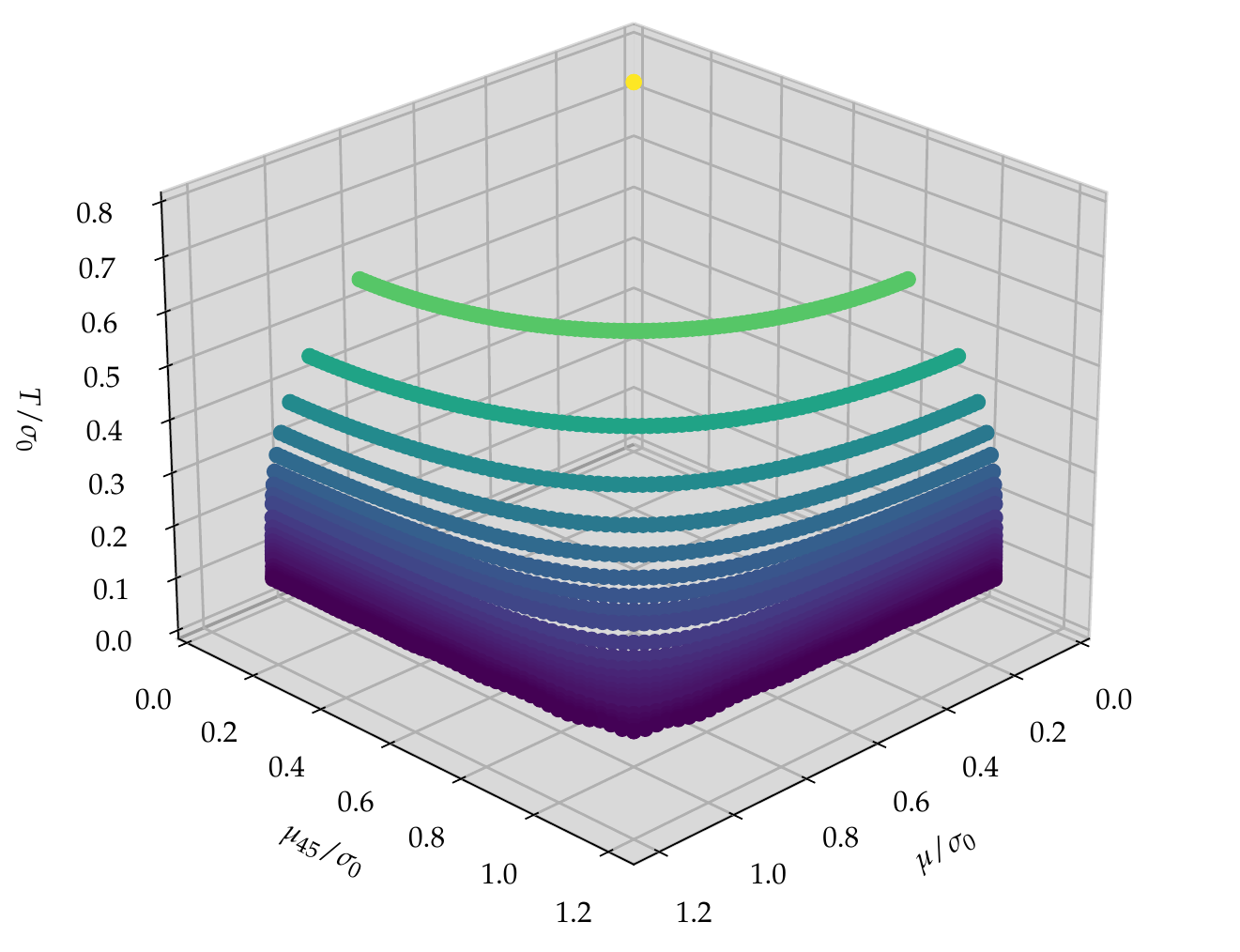}		
	\caption{\label{fig:iso2+1hom_pd}Phase diagram of the chirally imbalanced $2+1$-dimensional GN model with the restriction to a homogeneous condensate $\sigma = \bar{\sigma}$ in $(\mu,\muff,T)$ space for $a \sigma_0 = 0.2327$ and $L \sigma_0 = 120 \, a \sigma_0 = 27.92$.}
\end{figure}

Figure \ref{fig:iso2+1hom_pd} shows the phase diagram in $(\mu,\muff,T)$ space for $a \sigma_0 = 0.2327$ and $L \sigma_0 = 120 \, a \sigma_0 = 27.92$.
For $\muff = 0.0$ the phase boundary is quite similar to the analytically obtained continuum result \cite{Klimenko:1987gi} with slight deviations due to discretization and finite volume effects.
At high temperature 
$T/\sigma_0 \gtapprox 0.4$ the phase boundary exhibited an approximate rotational symmetry in the $\mu$-$\muff$ plane, i.e., it  was crudely described by $\mu^2 + \muff^2 \approx (\mu_c(T))^2$.
In contrast to that, at low temperature, 
the phase boundary approached a square-like shape in the $\mu$-$\muff$ plane.

\begin{figure}[H]
	\includegraphics[width=.49\columnwidth]{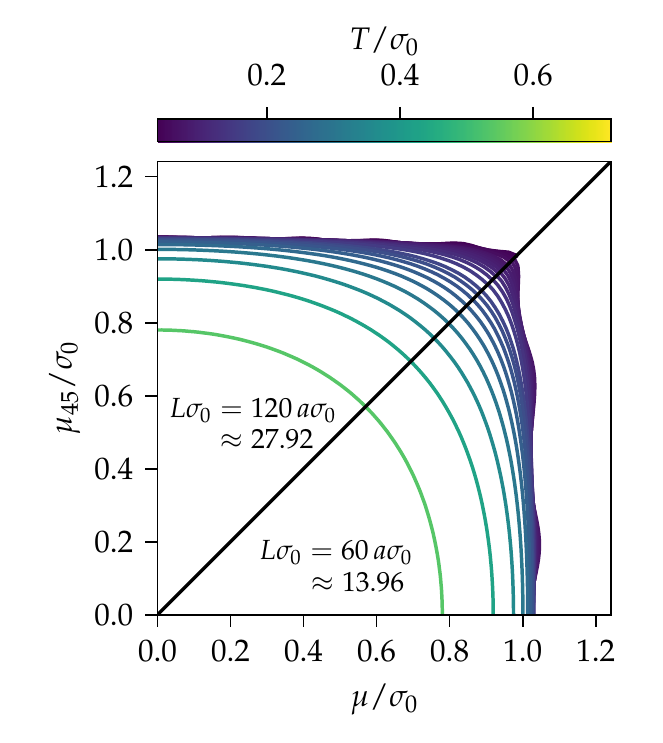}
	\includegraphics[width=.49\columnwidth]{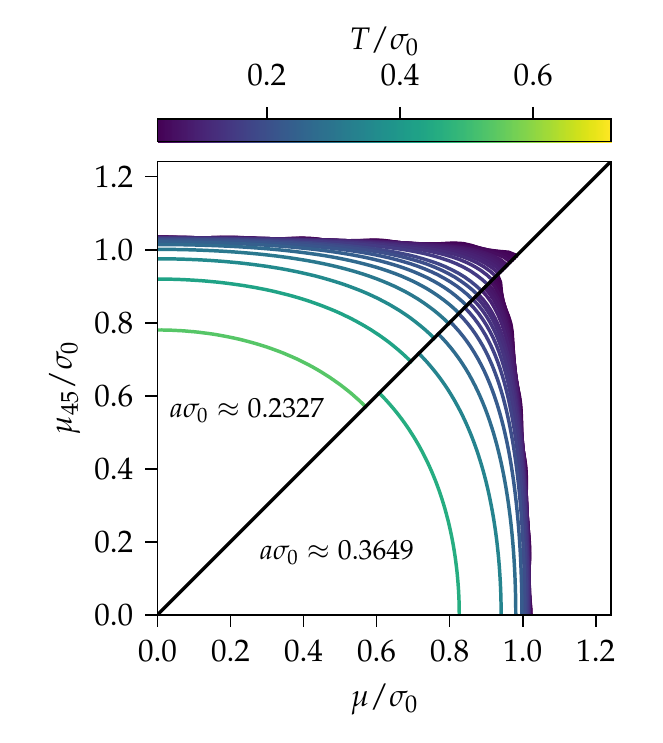}	
	\vspace{-6pt}	
	\caption{\label{fig:iso2+1hom_pd_2d}Phase diagram of the chirally imbalanced $2+1$-dimensional GN model with the restriction to a homogeneous condensate $\sigma = \bar{\sigma}$ in the $\mu$-$\muff$ plane for several temperatures. Since the phase diagram is invariant with respect to $\mu \leftrightarrow \muff$, each octant in the $\mu$-$\muff$ plane contains  full information and one can compare the two lattice extents $L \sigma_0$ (left plot) or two lattice spacings $a \sigma_0$ (right plot) in a convenient way within the same plot. (\textbf{left}) $a \sigma_0 = 0.2327$. (\textbf{right}) $\Ns/\Ntc = 20$, i.e., similar spatial lattice extents $L \sigma_0 = 80 \, a \sigma_0 = 29.19$ and $L \sigma_0 = 120 \, a \sigma_0 = 27.92$.}
\end{figure}

The left plot of Figure \ref{fig:iso2+1hom_pd_2d} shows sectional views of the phase diagram at fixed lattice spacing $a \sigma_0 = 0.2327$ for two different spatial extents, $L \sigma_0 = 60 \, a \sigma_0 = 13.95$  and $L \sigma_0 = 120 \, a \sigma_0 = 27.92$.
At low temperature the phase boundary exhibits an oscillatory behavior, which was more pronounced for the smaller lattice volume. We expected that the oscillations would disappear in the infinite volume limit.
The right plot of Figure \ref{fig:iso2+1hom_pd_2d} shows sectional views of the phase diagram for two different lattice spacings, $a \sigma_0 = 0.3649$ and $a \sigma_0 = 0.2327$, at fixed ratio $\Ns/\Ntc = 20$ implying similar spatial extents $L \sigma_0 = 80 \, a \sigma_0 = 29.19$ and $L \sigma_0 = 120 \, a \sigma_0 = 27.92$.
There were visible discrepancies due to discretization effects, in particular when both $T$ is small and $\mu \approx \muff$. Continuum results at $\muff=0$ from Ref.\ \cite{Klimenko:1987gi} as well as our lattice results at various small temperatures, lattice spacings and spatial volumes point towards $T = 0$ phase boundaries at $\mu/\sigma_0 = 1$ for $0 \leq \muff/\sigma_0 < 1$ and at $\muff/\sigma_0 = 1$ for $0 \leq \mu/\sigma_0 < 1$ in the continuum limit.

\vspace{-15pt}

\begin{figure}[H]
	\includegraphics[width=.49\columnwidth]{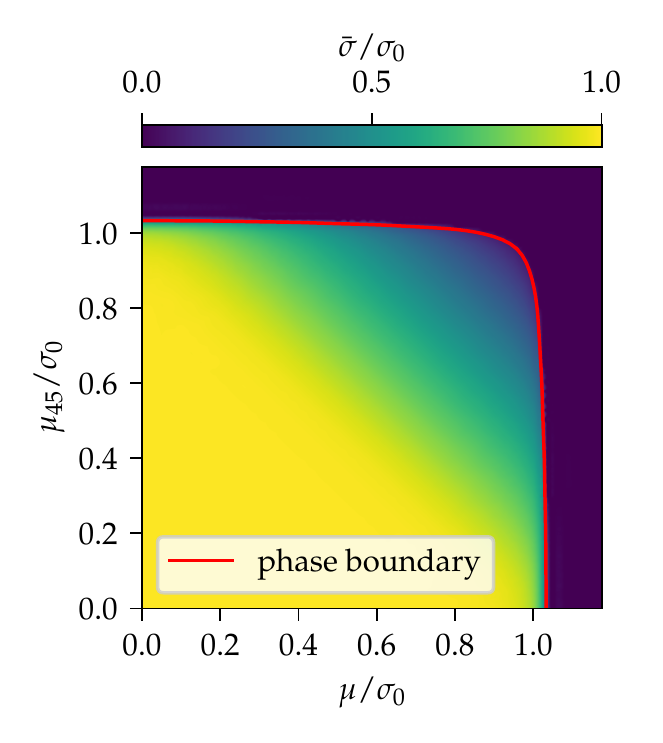}		
	\includegraphics[width=0.49\columnwidth]{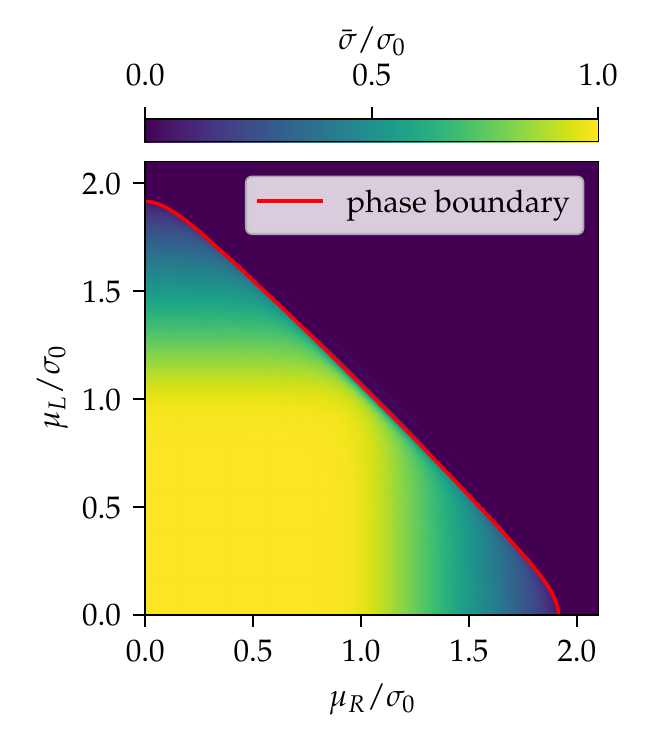}	
	\vspace{-9pt}	
	\caption{\label{fig:iso2+1hom_sigma}$\bar{\sigma} / \sigma_0$ for the chirally imbalanced $2+1$-dimensional GN model with the restriction to a homogeneous condensate $\sigma = \bar{\sigma}$ for $T/\sigma_0 = 0.0716$, $a \sigma_0 = 0.2327$ and $L \sigma_0 = 120 \, a \sigma_0 = 27.92$. (\textbf{left})~$\bar{\sigma} / \sigma_0$ as function of $\mu$ and $\muff$. (\textbf{right}) $\bar{\sigma} / \sigma_0$ as function of $\mu_L$ and $\mu_R$.}
\end{figure}

We also studied the behavior of $\bar{\sigma}$ at small temperature $T/\sigma_0 = 0.0716$, $a \sigma_0 = 0.2327$ and $L \sigma_0 = 120 \, a \sigma_0 = 27.92$. $\bar{\sigma}$ is shown as function of $\mu$ and $\muff$ in the left plot of Figure \ref{fig:iso2+1hom_sigma} and as function of $\mu_L = \mu + \muff$ and $\mu_R = \mu - \muff$ in the right plot of Figure \ref{fig:iso2+1hom_sigma}. To explain these results, we noted that the effective action \labelcref{eq:seff_LR} was the sum of a left-handed part $\seffL[\sigma]$ and a right-handed part $\seffR[\sigma]$ with chemical potentials $\mu_L$ and $\mu_R$, respectively. At the beginning of \cref{{results}}, we had already concluded that $\bar{\sigma} = 0$, if $|\mu_L| > \mu_c(T)$ and $|\mu_R| > \mu_c(T)$. Similarly, we argue now that both parts favor $\bar{\sigma} \approx \sigma_0$, if $|\mu_L| < \mu_c(T)$ and $|\mu_R| < \mu_c(T)$. The numerical results from Figure \ref{fig:iso2+1hom_sigma} are consistent with that expectation. In particular, the yellow regions, where $\bar{\sigma} \approx \sigma_0$, corresponds to $|\mu_L| < \mu_c(T)$ and $|\mu_R| < \mu_c(T)$. In the remaining regions of the $\mu$-$\muff$ plane, or equivalently the $\mu_L$-$\mu_R$ plane, $\seffL[\sigma]$ and $\seffR[\sigma]$ compete, leading to a continuous transition of the condensate from $\bar{\sigma} \approx \sigma_0$ to $\bar{\sigma} = 0$. This continuous behavior was consistent with the fact that the lattice GN model with the effective action $\seffL[\sigma]|_{\mu_L = \mu}$ or equivalently $\seffR[\sigma]|_{\mu_R = \mu}$, restricted to a homogeneous condensate, had a second-order phase transition at $T/\sigma_0 = 0.0716$, $a \sigma_0 = 0.2327$ and $L \sigma_0 = 120 \, a \sigma_0 = 27.92$. 

\vspace{-9pt}

We note that the lattice data shown in this subsection also represents a non-trivial cross-check of our implementation: all numerical results were consistent with the symmetry $\mu \leftrightarrow \muff$ within machine precision. 


\subsection{Stability of a Homogeneous Condensate \label{sec:instability}}

Now, we relaxed the constraint that $\sigma$ was a homogeneous condensate. To determine the preferred modulation of the condensate in a possibly existing inhomogeneous phase, we had to allow arbitrary spatial modulations of $\sigma$, i.e., consider $\sigma = \sigma(\x{x})$, and minimize the effective action with respect to these modulations. In lattice field theory this is possible but numerically very challenging. As a first step, therefore, we explored whether the homogeneous minima $\sigma = \bar{\sigma}$, which were determined in \cref{sec:hom_res} for many different $(\mu, \muff, T)$, were stable or unstable with respect to spatially inhomogeneous perturbations $\delta \sigma(\x{x})$. Boundaries between stable and unstable regions in $(\mu,\muff,T)$ space were identical to phase boundaries if the amplitude of the inhomogeneity became infinitesimal when approaching the boundary. However, a stability analysis failed to detect inhomogeneous condensates in regions of the phase diagram where the homogeneous minimum (found, e.g., as described in \cref{sec:hom_res}) corresponded to a local, but not global, minimum of $\seff[\sigma(\x{x})] / \Nf$. This was, e.g., the case in the $1+1$-dimensional GN model \cite{deForcrand:2006zz,Koenigstein:2021llr}.

A detailed derivation of the formalism to probe the stability of a homogeneous condensate $\sigma = \bar{\sigma}$ with respect to arbitrary spatial perturbations $\delta \sigma(\x{x})$ can be found in \Rcite{Buballa:2020nsi} for a continuum approach.
This formalism can be transferred to lattice discretizations in a straightforward way, which is discussed in the same reference. A quantity of central importance is
\begin{equation}
	\frac{\Gamma^{-1}(\x{q}_k)}{\Nf} = \frac{1}{\lambda} - \frac{\tilde{W}_2(\x{q}_k) \tilde{W}_2(-\x{q}_k)}{8} \sum_{p} \tr\left(\bar{Q}^{-1}_{p-q}[\mu,\muff,\bar{\sigma}] \bar{Q}^{-1}_p[\mu,\muff,\bar{\sigma}]\right) ,
\end{equation}
where $\sum_p$ runs over all 3-dimensional lattice momenta \labelcref{eq:3dMomenta}, $q = (0,\x{q}_k)$, the trace refers to spinor space and $\bar{Q}_\xt{p}[\mu,\muff,\bar{\sigma}]$ is defined via $\tilde{Q}^{(4)}_{\xt{p},\xt{q}}[\mu,\muff,\bar{\sigma}] = \kron{\xt{p}}{\xt{q}} \bar{Q}_\xt{p}[\mu,\muff,\bar{\sigma}]$ and \cref{eq:diracop_disc}, i.e.,
\begin{align}
	\label{eq:Qbar}
	\bar{Q}_\xt{p}[\mu,\muff,\bar{\sigma}] = \Nt\Ns^2 \left(\ii \sum_{\nu=0}^{2} \gamma_\nu \sin \Big(p_\nu - \kron{\nu}{0} \ii (\mu + \gamma_{45} \muff)\Big) + \bar{\sigma}\right) .
\end{align}

Negative values of $\Gamma^{-1}(\x{q}_k) / \Nf$ with $\x{q}_k \neq 0$ indicate instability of the condensate $\sigma = \bar{\sigma}$ with respect to harmonic perturbations with momentum $\x{q}_k$. Such perturbations decrease $\seff[\sigma]$; consequently, an inhomogeneous condensate was preferred.
By evaluating $\Gamma^{-1}(\x{q}_k) / \Nf$ for suitably chosen parameters $(\mu,\muff,T)$ we could identify regions that were part of an inhomogeneous phase.

\begin{figure}[H]
	\begin{subfigure}[b]{0.5\columnwidth}
		\includegraphics[width=\textwidth, trim= 0 1cm 0 1cm, clip]{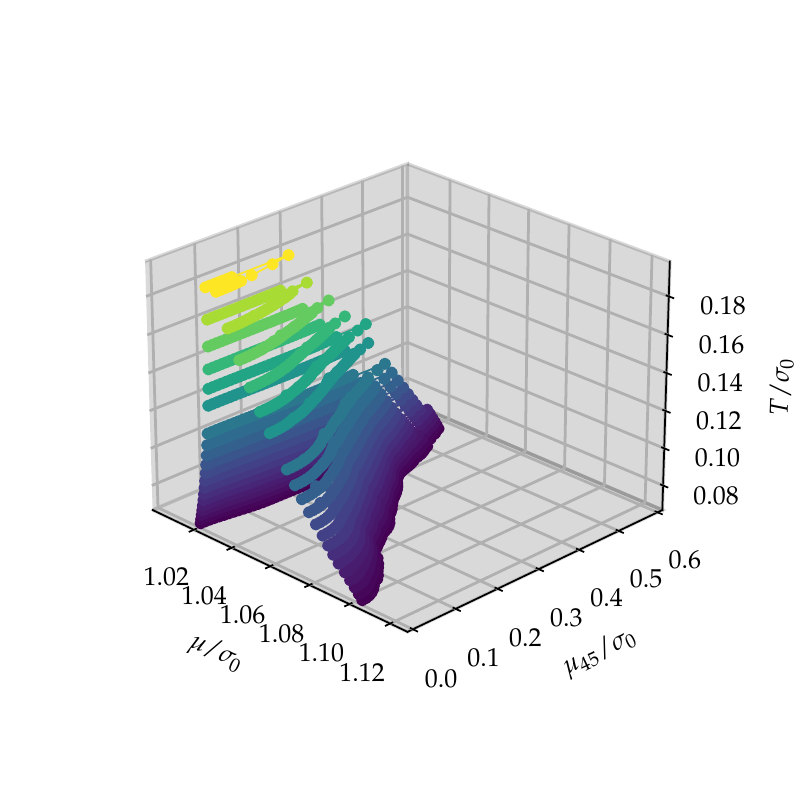}	
	\end{subfigure}
	\begin{subfigure}[b]{0.5\columnwidth}
		\includegraphics[width=\textwidth, trim= 0 1cm 0 1cm, clip]{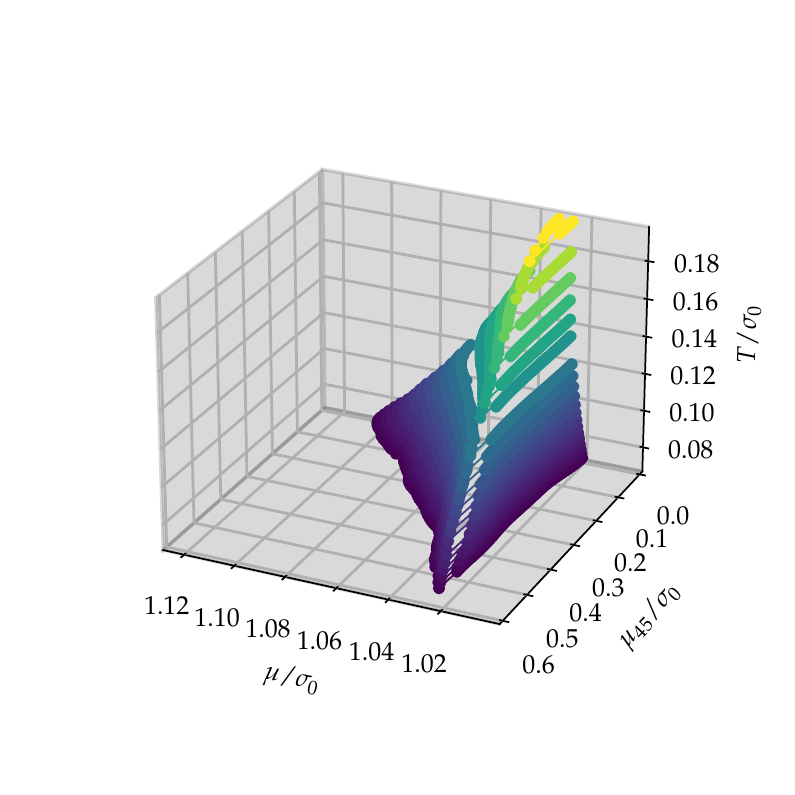}	
	\end{subfigure}
	\caption{\label{fig:iso_stab_analysis_3d}Boundaries	of the region of instability in the chirally imbalanced $2+1$-dimensional GN model for the discretization $\tilde{W}_2 = \tilde{W}''_2$ in $(\mu,\muff,T)$ space for $a \sigma_0 = 0.2327$ and $L \sigma_0 = 100 \, a \sigma_0 = 23.27$. Both plots show the same data and differ only in angle of view.}
\end{figure}

\begin{figure}[H]
	\includegraphics[width=.68\columnwidth]{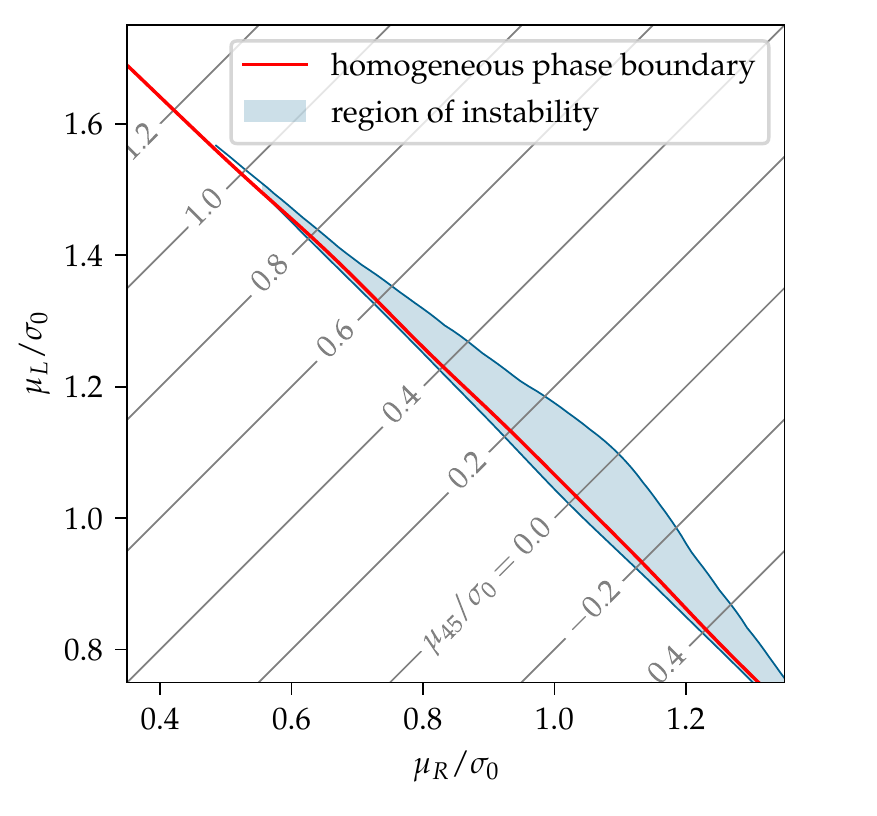}
	\vspace{-8pt}		
	\caption{\label{fig:iso_stab_analysis_2d_fixT}Region of instability and homogeneous phase boundary in the chirally imbalanced $2+1$-dimensional GN model for the discretization $\tilde{W}_2 = \tilde{W}'_2$ in the $\mu_R$-$\mu_L$ plane for $T/\sigma_0=0.076$, $a \sigma_0 = 0.2327$ and $L \sigma_0 = 100 \, a \sigma_0 = 23.27$. The chiral chemical potential $\muff$ was constant along the diagonal straight lines.}
\end{figure}

We searched extensively for regions, where $\bar{\sigma}$ was unstable, using both discretizations \labelcref{eq:W'_p} and \labelcref{eq:W''_p}. For $\tilde{W}_2 = \tilde{W}'_2$ such regions did not seem to exist. For $\tilde{W}_2 = \tilde{W}''_2$ and finite lattice spacing there was a region of instability at small $T$ consistent with the findings at $\muff = 0$ reported in \Rcite{Buballa:2020nsi}. Figure \ref{fig:iso_stab_analysis_3d} shows the lattice with the finer lattice spacing, $a \sigma_0 = 0.2327$, and spatial extent $L \sigma_0 = 100 \, a \sigma_0 = 23.27$. The region of instability was located within the tetrahedral shape. At smaller temperature it had a larger extent in the $\mu$-$\muff$ plane. A somewhat unexpected result was the large extent of the region of instability in $\muff$ direction (e.g., for $T/\sigma_0 = 0.076$ and $\mu/\sigma_0 \approx 1.0$ up to $\muff/\sigma_0 \approx 0.5$). Its boundary is plotted in Figure \ref{fig:iso_stab_analysis_2d_fixT} in the $\mu_R$-$\mu_L$ plane. The plot shows that the instability region extended up to $\muL = \mu + \muff \approx 1.5$ and at the same time down to $\muR = \mu - \muff \approx 0.5$. A symmetric phase was preferred by $\seffL[\sigma]$ with chemical potential $\muL \approx 1.5$, while $\seffR[\sigma]$ with chemical potential $\muR \approx 0.5$ prefered a homogeneous symmetry-broken phase. Thus, neither of the two parts of the effective action \labelcref{eq:seff_LR} favored an inhomogeneous phase, but in combination they did. This highlighted the non-trivial interplay of $\seffL[\sigma]$ and $\seffR[\sigma]$ gave rise to a rather large inhomogeneous phase at finite lattice spacing for certain discretizations. Note that Figure \ref{fig:iso_stab_analysis_2d_fixT} also reveals that the homogeneous phase boundary was engulfed by the region of instability, which was not the case in our previous study at $\muff = 0$ \cite{Buballa:2020nsi}, where a different lattice regularization was used.

\vspace{-6pt}

\begin{figure}[H]
	\includegraphics[width=0.95\columnwidth]{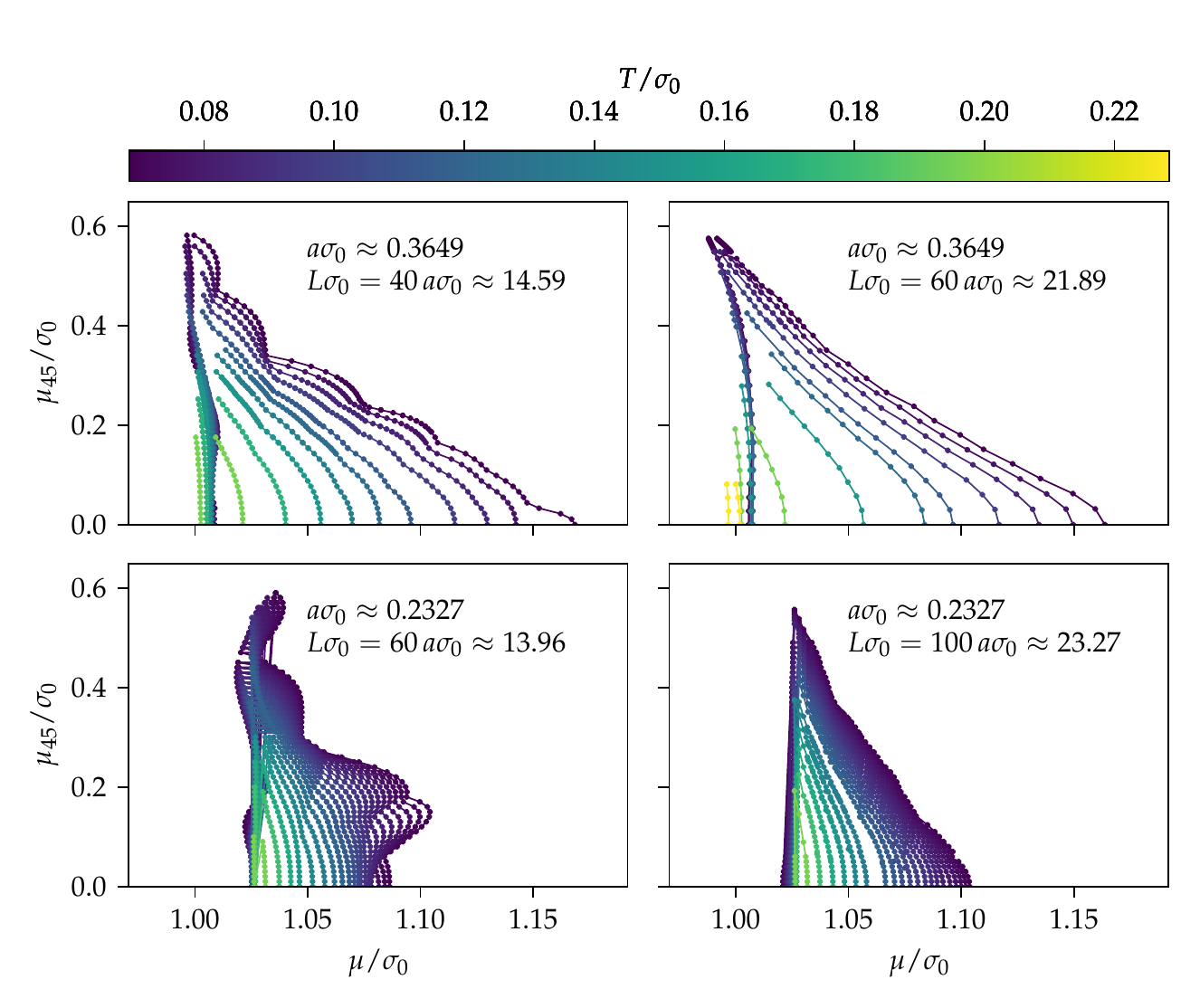}	
	\vspace{-18pt}	
	\caption{\label{fig:iso_stab_analysis_2d}Boundaries of the region of instability in the chirally imbalanced $2+1$-dimensional GN model for the discretization $\tilde{W}_2 = \tilde{W}''_2$ in the $\mu$-$\muff$ plane for several temperatures, two different $a \sigma_0$ (upper versus lower row) and two different $L \sigma_0$ (left versus right column).}
\end{figure}

In Figure \ref{fig:iso_stab_analysis_2d} we show sectional views of the region of instability for the discretization $\tilde{W}_2 = \tilde{W}''_2$ and various temperatures. The upper row corresponds to the larger lattice spacing $a \sigma_0 = 0.3649$ and the lower row to the smaller lattice spacing, $a \sigma_0 = 0.2327$, while the left column corresponds to smaller and the right column to larger spatial extent $L \sigma_0$.
From comparing the upper and the lower row, it is obvious that the instability region shrank when  the lattice spacing decreased, most prominently in the $\mu$ direction.
This is particularly evident in the right column, where the boundaries are significantly less distorted by finite volume effects.
In the plots in the left column, however, there are pronounced oscillations that seem to be caused by small spatial volume.
These oscillations are reminiscent of those observed in the $\mu$-$T$ plane in lattice studies of the chirally balanced GN model in $1+1$ and $2+1$ dimensions \cite{deForcrand:2006zz,Winstel:2019zfn,Buballa:2020nsi}.

In summary, we found no region of instability for $\tilde{W}_2 = \tilde{W}'_2$ but a shrinking region of instability for decreasing lattice spacing for $\tilde{W}_2 = \tilde{W}''_2$. This {strongly suggests} that there is no region of instability in the chirally imbalanced $2+1$-dimensional GN model in the continuum limit.

\subsection{Arbitrary Spatial Modulations of the Condensate \label{sec:min}}

Now, we discuss the minimization of the effective action \labelcref{eq:seff_momentum_disc} with respect to the condensate allowing arbitrary spatial modulations, i.e.,\ arbitrary Fourier coefficients $\tilde{\sigma}(\x{p})$.
We did this for selected parameters $(\mu,\muff,T)$ by carrying out several conjugate gradient minimizations, which differed in starting values for $\tilde{\sigma}(\x{p})$.
For each $(\mu,\muff,T)$ we found only a small number of local minima although a significantly larger number of different starting values for $\tilde{\sigma}(\x{p})$ were provided to the minimization algorithm.
This might have indicated that for all considered $(\mu,\muff,T)$ the corresponding global minimum was among the found local minima.
We note that in our previous work \cite{Buballa:2020nsi} only 1-dimensional modulations were studied, i.e., $\tilde{\sigma} = \tilde{\sigma}(p_1)$.
In this work we relaxed that constraint to allow arbitrary 2-dimensional modulations, i.e., $\tilde{\sigma} = \tilde{\sigma}(\x{p})$.
Since this was a numerically difficult and computer time-intensive task, we used a rather small lattice with coarse lattice spacing $a \sigma_0 = 0.3649$, temperature $T/\sigma_0 = 0.114$, $L \sigma_0 = 28 \, a \sigma_0 = 10.22$ and discretization corresponding to $\tilde{W}_2 = \tilde{W}''_2$.

\begin{figure}[H]
	\begin{center}
		{\small 1-dimensional modulations corresponding to global minima, $\muff = 0$}
		\includegraphics{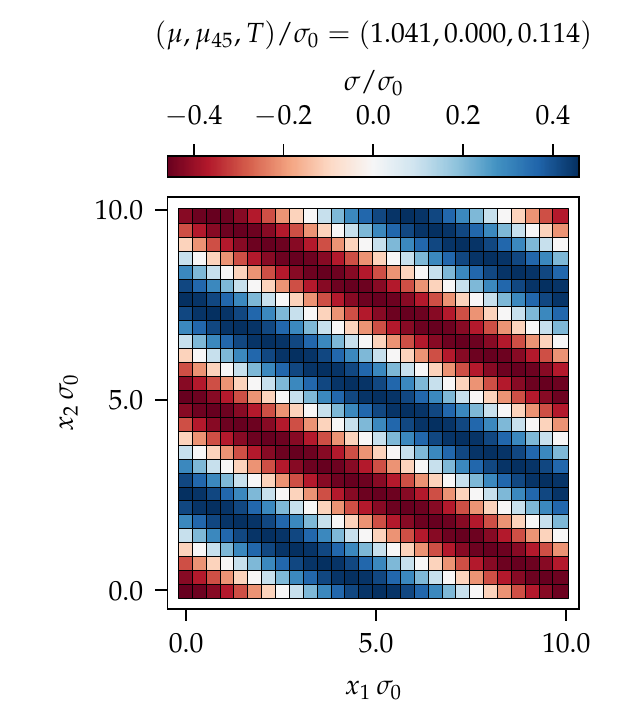}
		\includegraphics{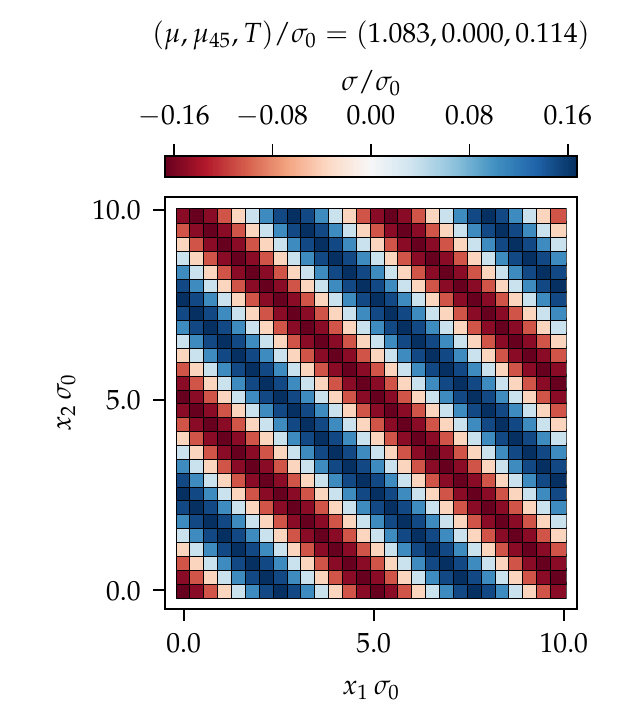}	\\
		\vspace{-3pt}
		{\small 2-dimensional modulation corresponding to a local minimum, $\muff = 0$}
		\includegraphics{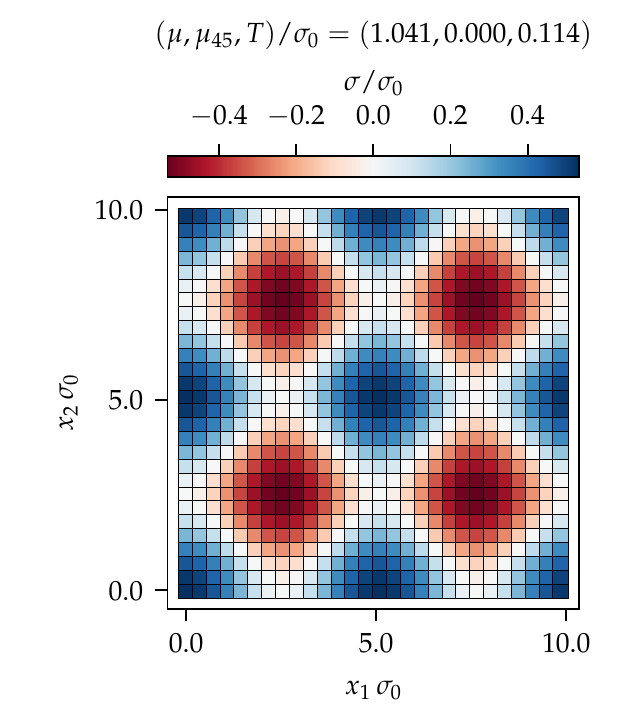}	\\
		\vspace{-3pt}
		{\small 1-dimensional modulations corresponding to global minima, $\muff \neq 0$}
		\includegraphics[trim= 0 6 0 0, clip]{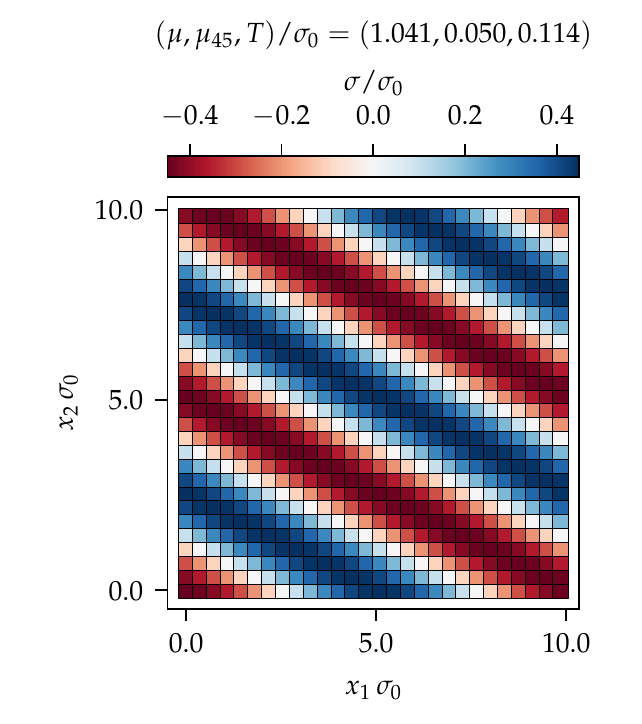}	 
		\includegraphics[trim= 0 6 0 0, clip]{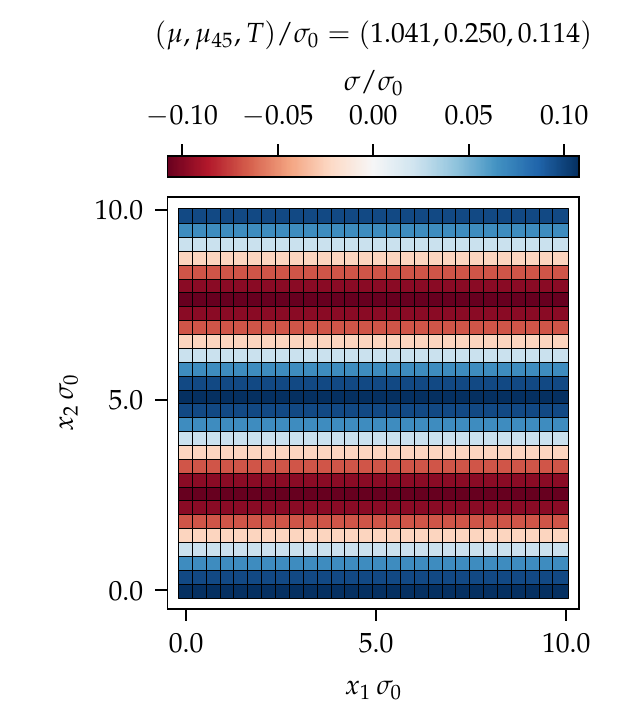}	
	\end{center}
	\vspace{-9pt}
	\caption{\label{fig:2d_minima}Modulations of the condensate $\sigma$ corresponding to minima of the effective action \labelcref{eq:seff_momentum_disc} for the discretization $\tilde{W}_2 = \tilde{W}''_2$, $T/\sigma_0 = 0.114$, $a \sigma_0 = 0.3649$ and $L \sigma_0 = 28 \, a \sigma_0 = 10.22$.}
\end{figure}

First, we studied the chirally balanced model, i.e.,\ $\muff = 0$.
As an example, the upper left plot in Figure \ref{fig:2d_minima} shows a configuration $\sigma(\x{x})$ corresponding to one of the global minima of the effective action at $\mu/\sigma_0 = 1.041$.
An inhomogeneous condensate was favored, as we learned 
from the stability analysis discussed in \cref{sec:instability}.
Even though the minimization algorithm allows arbitrary $2$-dimensional modulations, the resulting global minimum was just a plane wave with wave vector $\x{q}_k = 2 \pi (1,2) / L$.
Similarly, the upper-right plot in Figure \ref{fig:2d_minima} shows a minimizing condensate at a larger chemical potential $\mu/\sigma_0 = 1.083$.
Again, we found a plane wave, but this time with wave vector $\x{q}_k = 2 \pi (2,2) / L$, i.e., with a smaller wavelength.
Even though the found plane waves were $1$-dimensional structures, by allowing arbitrary $2$-dimensional modulations it reduced finite volume corrections. This was so because, in contrast to our previous work \cite{Buballa:2020nsi}, the direction of the wave vector was no longer restricted to being parallel to one of the coordinate axes; thus, its magnitude could be changed in finer steps.

The minimization algorithm also found 2-dimensional modulations.
These, however, corresponded exclusively to local minima of the effective action \labelcref{eq:seff_momentum_disc}.
An example is shown in the center of Figure \ref{fig:2d_minima}.

We also investigated, how chiral imbalance, i.e.,\ $\muff \neq 0$, affected the preferred modulation of the condensate.
The plots in the lower row of Figure \ref{fig:2d_minima} show the global minima of the effective action for $(\mu/\sigma_0,\muff/\sigma_0) = (1.041,0.05)$ and $(\mu/\sigma_0,\muff/\sigma_0) = (1.041,0.25)$.
At fixed $\mu/\sigma_0 = 1.041$ the frequency was almost independent of $\muff$ (cf.\ the upper left plot, the lower left plot and the lower right plot of Figure \ref{fig:2d_minima}). The amplitude, however, decreased when increasing $|\muff|$, suppored a second-order phase transition, also at $\muff \neq 0$.
We searched extensively for inhomogeneous condensates outside the instability region explored and discussed in \cref{sec:instability}, but did not find any.
Thus, we concluded that the boundaries of the instability region were identical to phase boundaries and that the corresponding phase transitions were of the second order.

\section{Conclusions\label{conclusion}}

In this work we studied the phase diagram of the $2+1$-dimensional GN model with chiral imbalance introduced via a chiral chemical potential $\muff$ {using the mean-field approximation}. 
Our lattice field theory results indicated that an inhomogeneous phase exists at finite lattice spacing $a$, when using a specific lattice discretization ({$\tilde{W}_2 = \tilde{W}''_2$}). Non-vanishing $\muff$, however, seems to disfavor inhomogeneous modulations. Moreover, the inhomogeneous phase shrank for decreasing $a$ and was expected to disappear in the continuum limit. These findings are consistent with our previous work \cite{Buballa:2020nsi}, which was restricted to $\muff = 0$. (For completeness we note that inhomogeneous condensates at $T = 0$ and $\muff = 0$, which are energetically degenerate to the homogenous condensate in the homogeneous symmetry-broken phase, were found in \Rcite{Urlichs:2007zz}. Our numerical lattice studies were, however, limited to $T > 0$).
{Investigations of $1+1$-dimensional GN-type models \cite{Cohen:1983nr,Karsch:1986hm,Lenz:2020bxk, Lenz:2020cuv, Lenz:2021kzo, Stoll:2021ori, Lenz:2021vdz, Horie:2021wnn} at finite $\Nf$ showed that bosonic quantum fluctuations can influence the extent and existence of (in-)homogeneous phases significantly. Such fluctuations are likely to affect the extent of the homogeneous symmetry-broken phase in the present $2+1$-dimensional GN model. The inhomogeneous phase was expected to remain absent since bosonic quantum fluctuations tend to disfavor ordered phases even more.}

Moreover, for our chirally imbalanced $2+1$-dimensional GN model, we showed that an isospin chemical potential $\muI$ is equivalent to the chiral chemical potential $\muff$. Thus, all results presented can either be interpreted in the context of chiral imbalance or of isospin imbalance. In particular the $\mu$-$\muff$-$T$ phase diagram was identical to the $\mu$-$\mu_I$-$T$ phase diagram. Interestingly, a recent study of the $3+1$-dimensional NJL model in the large-$N_c$ limit \cite{Khunjua:2018dbm} conjectured a similar approximate duality of the phase diagram.

{Color-superconductivity might play an important role in such studies \cite{Sadzikowski:2006jq, lakaschus:2020caq}, especially at finite isospin chemical potential \cite{Fukushima:2007bj, Nowakowski:2016dwu}}. In the considered $2+1$-dimensional GN model this could, however, not be investigated, because the necessary difermion interaction was not present. Thus, we are not yet in a position to compare our results to up-to-date lattice QCD simulations at finite $\muI$ (see, e.g., \Rcite {Brandt:2016zdy,Brandt:2019hel,Brandt:2019ttv,Cao:2020ske,Brandt:2021yhc}), where a phase with Bose--Einstein condensation of charged pions was observed. As a next step, it might therefore be interesting to establish contact with \Rcite{Ebert:2016ygm,Zhukovsky:2017hzo}, where a color-superconducting channel was added to the chirally imbalanced $2+1$-dimensional GN model. It should be noted, however, that these references introduced the chiral chemical potential in a conceptually different way using spin matrices $\gamma_0 \gamma_4$ as well as $\gamma_0 \gamma_5$, instead of $\gamma_0 \gamma_{45}$, as in Equation\ (\ref{eq:block_Q4}).


\authorcontributions{All authors contributed equally to this work. All authors have read and agreed to the published version of the manuscript.}

\funding{
	L.P.,  M.Wa.\ and M.Wi.\ acknowledge support by the Deutsche Forschungsgemeinschaft (DFG, German Research Foundation) through the CRC-TR 211 ``Strong-interaction matter under extreme conditions''---project number 315477589---TRR 211.
	L.P. and M.W.\ acknowledge the support of the Helmholtz Graduate School for Hadron and Ion Research.
	M.Wa.\ acknowledges support by the Heisenberg Programme of the Deutsche Forschungsgemeinschaft (DFG, German Research Foundation)---project number 399217702.
	M.Wi.~acknowledges support by the GSI Forschungs- und Entwicklungsvereinbarungen (GSI F\&E). 
	M.Wi.~acknowledges the support of the Giersch Foundation.
}

\institutionalreview{Not applicable}

\informedconsent{Not applicable}

\dataavailability{The data set is included in the source files of the pre-print uploaded on \url{https://arxiv.org/abs/2112.11183} (accessed on 21 December 2021).} 

\acknowledgments{We would like to thank M.~Buballa, A.~Königstein, L.~Kurth, A.~Sciarra and M.~Thies for valuable discussions regarding this work. We further thank J.~Lenz, M.~Mandl, R.D.~Pisarksi and A.~Wipf for general valuable discussions regarding inhomogeneous phases and four-fermion theories. Numerical results in this work were produced with \texttt{C++} code using the \texttt{nlopt} library \cite{Johnson2011} for numerical minimization and the \texttt{eigen3} library for matrix operations \cite{eigenweb}. The data used in the figures were processed with the \texttt{pandas} \cite{reback2020pandas} and \texttt{numpy} \cite{2020NumPy-Array} packages. All figures in this work were produced with the \texttt{matplotlib} package \cite{Hunter:2007}. Calculations on the GOETHE-HLR and on the FUCHS-CSC high-performance computer of the Frankfurt University were conducted for this research. We would like to thank HPC-Hessen, funded by the State Ministry of Higher Education, Research and the Arts, for programming advice.}

\conflictsofinterest{The authors declare no conflict of interest.} 

\begin{adjustwidth}{-\extralength}{0cm}

\reftitle{References}


\end{adjustwidth}
\end{document}